\documentclass{Interspeech2024}

\usepackage{cite}
\usepackage[acronym]{glossaries}

\newacronym{bam}    {BAM}       {binary activation map}

\newacronym{cnn}    {CNN}       {convolutional neural network}

\newacronym{dns}    {DNS}       {deep noise suppression}

\newacronym{mse}    {MSE}       {mean squared error}
\newacronym{mos}    {MOS}       {mean opinion score}

\newacronym{pcc}    {PCC}       {Pearson correlation coefficient}
\newacronym{pesq}   {PESQ}      {perceptual evaluation of speech quality}
\newacronym{ptq}    {PTQ}       {post-training quantization}

\newacronym{qat}    {QAT}       {quantization aware training}

\newacronym{si-sdr} {SI-SDR}    {scale-invariant signal-to-distortion ratio}
\newacronym{snr}    {SNR}       {signal-to-noise ratio}
\newacronym{sqp}    {SQP}       {speech quality prediction}
\newacronym{stft}   {STFT}      {short-time Fourier transform}
\newacronym{stoi}   {STOI}      {short-time objective intelligibility}
 
\interspeechcameraready  

\title{Resource-Efficient Speech Quality Prediction
    through\\Quantization Aware Training
    and Binary Activation Maps}

\name[affiliation={1,*}]{Mattias}{Nilsson}
\name[affiliation={2,3,*}]{Riccardo}{Miccini}
\name[affiliation={2}]{Clément}{Laroche}
\name[affiliation={2}]{Tobias}{Piechowiak}
\name[affiliation={1,4}]{Friedemann}{Zenke}

\address{\normalsize
    $^1$Friedrich Miescher Institute for Biomedical Research, Switzerland,
    $^2$GN Audio, Denmark,\\
    $^3$Technical University of Denmark, Denmark,
    $^4$University of Basel, Switzerland,
    $^*$equal contributions}

\email{\normalsize
    mattias.nilsson@fmi.ch,
    rimi@dtu.dk,
    claroche@jabra.com,
    topiechowiak@jabra.com,
    friedemann.zenke@fmi.ch
}


\begin{document}

\maketitle

\begin{abstract}
As speech processing systems in mobile and edge devices become more commonplace, the demand for unintrusive speech quality monitoring increases.
Deep learning methods provide high-quality estimates of objective and subjective speech quality metrics.
However, their significant computational requirements are often prohibitive on resource-constrained devices.
To address this issue, we investigated binary activation maps (BAMs) for speech quality prediction on a convolutional architecture based on DNSMOS.
We show that the binary activation model with quantization aware training matches the predictive performance of the baseline model.
It further allows using other compression techniques.
Combined with 8-bit weight quantization, our approach results in a 25-fold memory reduction during inference, while replacing almost all dot products with summations.
Our findings show a path toward substantial resource savings by supporting mixed-precision binary multiplication in hard- and software. \end{abstract}

\section{Introduction}
Power-constrained audio devices such as headsets, earbuds, and hearing aids are driving the need for resource-efficient and real-time speech enhancement solutions.
Denoising based on deep learning---so-called \gls{dns}---is surpassing conventional speech enhancement methods based on signal processing by its ability to handle complex real-world noise \cite{tzinis_sudo_2020, bralios_latent_2023, fedorov_tinylstms_2020}.
When evaluating \gls{dns} systems, commonly used metrics include objective ones, such as \gls{pesq} \cite{rix_perceptual_2001}, \gls{stoi} \cite{taal_algorithm_2011}, and \gls{si-sdr} \cite{roux_sdr_2019}, as well as subjective \glspl{mos}.
Each of these metrics features both strengths and weaknesses.
Specifically, objective metrics are computationally efficient but do not always correlate well with human perception \cite{reddy_scalable_2019} and require a clean reference signal, which makes them difficult to use with real-world data.
Conversely, subjective metrics are more accurate and reliable, but require human listeners and are therefore more expensive and time-consuming to obtain.

In recent years, \gls{sqp} has emerged as a promising way to overcome the limitations of existing quality metrics by using machine learning.
\Gls{sqp} systems can be trained to predict both objective and subjective metrics directly from a noisy or processed speech signal, without the need for clean reference signals or human listeners, thereby making such systems ``unintrusive", i.e.\ reference-less.
Notable recent works on unintrusively estimating objective metrics target \gls{pesq} \cite{fu_quality-net_2018, jayesh_transformer_2022, yu_metricnet_2021} and \gls{stoi} \cite{zezario_stoi-net_2020}, as well as multiple metrics simultaneously \cite{kumar_torchaudio-squim_2023, catellier_wideband_2023, zezario_deep_2023}.
Similarly, \gls{sqp} systems aimed at estimating user \gls{mos} have also been developed \cite{reddy_dnsmos_2021, mittag_nisqa_2021, manocha_noresqa_2021}.

To make real-time \gls{dns} more efficient, common approaches include
architectural adjustments such as the use of depth-wise separable convolutions \cite{tzinis_sudo_2020},
model compression techniques such as quantization \cite{fedorov_tinylstms_2020},
as well as aspects of dynamic neural networks such as
slimmable channels \cite{elminshawi_slim-tasnet_2023},
layer skipping \cite{bralios_latent_2023},
or early exiting \cite{miccini_dynamic_2023, chen_dont_2021, li_learning_2021, sun_progressive_2020}.
This latter class of models can adapt its computational graph at inference time depending on the input, which is particularly useful in real-world applications with substantially changing noise conditions.
To accomplish this, dynamic neural networks rely on exiting or skipping policies that are often hand-crafted using simple heuristics \cite{chen_dont_2021, li_learning_2021}, or small sub-networks that are trained end-to-end with the \gls{dns} system \cite{bralios_latent_2023}, often using differentiable relaxation of a discrete distribution \cite{bengio_estimating_2013} which is difficult to train.
Importantly, none of the above strategies take speech quality into account.

While some attempts have been made at estimating concrete characteristics of the speech signal with the purpose of affecting the routing of a downstream model \cite{sun_progressive_2020}, this area of research is little explored---arguably due to the computational demands currently associated with even the smallest \gls{sqp} model.
Here, we argue that an adequately efficient and lean \gls{sqp} system will benefit dynamic noise suppression, allowing for continuous inference in real time.
Furthermore, the aforementioned ubiquity of \gls{dns} systems, especially on embedded and wearable devices, which might experience significantly different conditions than originally trained for, calls for tighter scrutiny and monitoring of performance to avoid further signal degradation in the form of unwanted artifacts \cite{ho_naaloss_2023}.

Current \gls{sqp} solutions range from having tens of thousands of trainable parameters \cite{reddy_dnsmos_2021} to several millions \cite{kumar_torchaudio-squim_2023}.
However, due to the long sequences used for prediction, storing the activations of the early layers remains demanding, why reducing their memory footprint through quantization is a primary goal.
While \gls{ptq} schemes exist, they perform poorly for binarization---an extreme form of quantization.
Still, binarization is particularly appealing in the \gls{sqp} setting because it minimizes memory consumption while also allowing for replacement of many algorithmic computations with bitwise operations.
To reach good task performance with binarized neural networks, \gls{qat} is imperative.
Standard \gls{qat} techniques for binarized neural networks are surrogate gradients, which are nonlinear variants of straight-through estimators \cite{bengio_estimating_2013, courbariaux_binarized_2016, neftci_surrogate_2019}.

In this work, we investigated reduction of the computational cost of \gls{sqp}, thereby paving the way for speech-quality driven dynamic denoising.
Specifically, we examined the computational benefits of \glspl{bam} and weight quantization in the DNSMOS architecture \cite{reddy_dnsmos_2021}---a \gls{cnn} commonly used for evaluating denoising systems.
For binarizing the activations, we used a \gls{qat} method based on a straight-through estimator \cite{bengio_estimating_2013} with a nonlinear surrogate derivative commonly used for optimizing spiking neural networks \cite{neftci_surrogate_2019}.
We show that \glspl{bam} combined with non-binary weights result in substantial resource savings while hardly affecting the performance of the model in terms of common evaluation metrics. 
\section{Methods}

\subsection{Problem Definition}
\label{sec:problem}

Given a reference speech signal $s$, we denote its degraded analogue as $s_d$.
Conventional intrusive algorithms, such as \gls{pesq} \cite{rix_perceptual_2001}, compute a speech quality metric $\textsc{sq}$ using both signals:
\begin{equation}
    \textsc{sq} = f(s, s_d).
\end{equation}
In the context of \gls{sqp}, we aim to train a deep learning model to estimate the metric using only the degraded signal:
\begin{equation}
    \operatorname*{arg\,min}_{\theta \in \mathbb{R}} \mathcal{L}(\textsc{sq}, \hat{\textsc{sq}});
    \quad \hat{\textsc{sq}} = f_\theta(s_d),
\end{equation}
where $\hat{\textsc{sq}}$ is the estimated speech quality metric, $\mathcal{L}$ is a cost or loss function, and $\theta$ represents the model parameters.

\subsection{Quantization}
\label{sec:quantization}

In this work, we compared different quantization approaches.
Specifically, we looked at binary activations, binary weights, and 8-bit quantization of weights and activations.

\subsubsection{Binary Activation Maps (BAMs)}

To binarize the activation maps of DNSMOS, we replaced the activations of the convolutional layers of the model with the Heaviside step function:
\begin{equation}
    H(x) = 
    \begin{cases} 
        0 & \text{for } x < 0 \\
        1 & \text{for } x \geq 0 \quad.
    \end{cases}
\end{equation}
This makes the activations discontinuous and, therefore, non-differentiable.
Note that we refer to networks with \glspl{bam} as binarized networks, even when their weights are non-binary.
In addition to the change in activation function in the convolutional layers, we used global average pooling instead of global max pooling after the final convolutional layer in all binarized networks.
In Sec.~\ref{sec:surrogate_gradient}, we present a \gls{qat} method that we used to circumvent the problem of non-differentiable activations.

\subsubsection{Binary Weights}
\label{sec:binary_weights}

In simulations with binary weights, we wrapped the floating-point weight parameters of the convolutional kernels with a scaled and shifted step function:
\begin{equation}
    Q(w) = 
    \begin{cases} 
        -1 & \text{for } w < 0 \\
        1 & \text{for } w \geq 0 \quad .
    \end{cases}
\end{equation}
This shift in baseline was done to ensure zero mean preactivations.

\subsubsection{8-Bit Quantization}

The 8-bit quantization was done in the form of static \gls{ptq} using in-built functionality in PyTorch.
This process involves a calibration step in which batches of data are fed through the model and observer modules are used to record the resulting distributions of activations and weights.
As input data for the calibration, we used a subset of 20\,\% of the training data.
We used the default \gls{ptq} configuration in PyTorch for the ``x86" backend.
During calibration, activations are divided into 2,048 dynamic histogram bins, which are then subject to a search for the optimal minimum and maximum values that minimize the quantization error.
Conversely, weights are quantized by recording their maximum and minimum values on a per-channel basis.
These minima and maxima are then employed to derive the scaling and zero-point coefficients of the affine quantization transform \cite{jacob_quantization_2018}.

\subsection{Surrogate Gradient}
\label{sec:surrogate_gradient}

\begin{figure}[tbh]
    \centering
    \includegraphics[width=0.8\columnwidth]{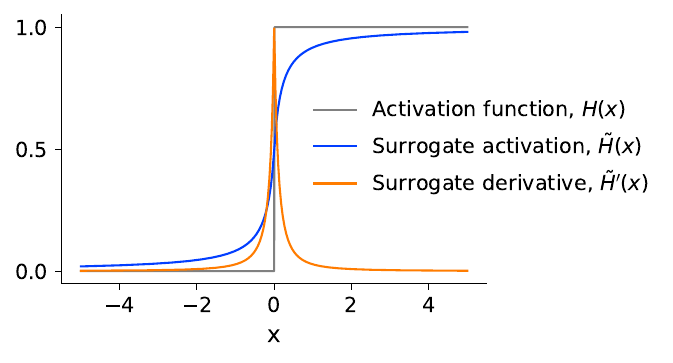}
    \caption{\textbf{Surrogate derivative for training of neural networks with binary activations.}
The surrogate activation function is a fast sigmoid with adjustable steepness.
    }
    \label{fig:surr_grad}
\end{figure}

To circumvent the problem of non-differentiability during \gls{qat} of the binarized networks, we used a surrogate gradient approach \cite{neftci_surrogate_2019}, which extends on the straight-through estimator proposed in \cite{bengio_estimating_2013}.
To that end, we replaced the ill-defined derivatives of the discontinuous Heaviside activation function that appear in the backward pass with the derivatives of continuous surrogate activation functions (Fig.~\ref{fig:surr_grad}).
Specifically, we used the SuperSpike surrogate derivative \cite{zenke_superspike_2018}, which corresponds to the normalized derivative of a fast sigmoid:
\begin{equation}
    \label{eq:SuperSpike}
    \tilde{H}'(x) = \frac{1}{\left(\beta\lvert x \rvert + 1\right)^2},
\end{equation}
where $\beta$ is a steepness parameter.

The above surrogate gradient was used as a drop-in replacement for a regular gradient during the backward pass, and, otherwise, training proceeded through standard back-propagation. 
\section{Experimental Setup}
We applied and present our findings on the DNSMOS model \cite{reddy_dnsmos_2021}, which is relatively compact and often used to unintrusively estimate \gls{mos} on speech enhancement works \cite{dubey_icassp_2024, timcheck_intel_2023}.
The code used to conduct our experiments is available at \url{https://github.com/fmi-basel/binary-activation-maps-sqp}.

\subsection{Model}
\label{sec:model}

The model consists of a \gls{cnn} taking in a 2-dimensional representation of an audio signal in the form of a log-power Mel-scale spectrogram.
The input is mapped into progressively more compact representations through convolutional and pooling layers.
The resulting features are then averaged along the spectral and temporal dimensions and further processed by a series of fully connected layers to derive a real-valued scalar estimation of a \gls{mos}.
Compared with the original DNSMOS architecture, we used a wider \gls{stft} window of 40\,ms (while maintaining the 50\,\% overlap) to ensure correct processing of Mel filter banks.
The model and its parameters are presented in Table~\ref{tab:dnsmos}.
For an exhaustive description of the architecture and preprocessing steps, see \cite{reddy_dnsmos_2021}.

\begin{table*}[th]
    \centering
    \caption{\textbf{Baseline model for speech quality prediction, DNSMOS.}
The numbers of parameters and multiply--add operations were retrieved using torchinfo \cite{Yep_torchinfo_2020}.
    }
    \footnotesize
\begin{tabular}{lrrrr}
        \toprule
        \textbf{Layer}                  & \textbf{Output shape} & \textbf{Parameters}   & \textbf{Multiply--adds} & \textbf{Activations} \\
        \midrule
        Input                           & 449x120x1 \\             
        \midrule
        Conv2D-1 (3x3), ReLU            & 449x120x32            & 320                   & 17,241,600            & 1,724,160 \\
        MaxPool2D-1 (2x2), Dropout(0.3) & 224x60x32 \\
        \midrule
        Conv2D-2 (3x3), ReLU            & 224x60x32             & 9,248                 & 124,293,120           & 430,080 \\
        MaxPool2D-2 (2x2), Dropout(0.3) & 112x30x32 \\
        \midrule
        Conv2D-3 (3x3), ReLU            & 112x30x32             & 9,248                 & 31,073,280            & 107,520 \\
        MaxPool2D-3 (2x2), Dropout(0.3) & 56x15x32  \\
        \midrule
        Conv2D-4 (3x3), ReLU            & 56x15x64              & 18,496                & 15,536,640            & 53,760 \\
        GlobalMaxPool2D$^1$                 & 1x64      \\
        \midrule
        Dense-1                         & 1x64                  & 4,160                 & 4,160                 & 64 \\
        Dense-2                         & 1x64                  & 4,160                 & 4,160                 & 64 \\
        Dense-3                         & 1x1                   & 65                    & 65                    & 1 \\
        \midrule
        Total                           &                       & 45,697                & 188,153,025           & 2,315,649 \\
        \bottomrule
    \end{tabular}
    
    {\scriptsize $^1$ We used global average pooling in the binarized networks instead.}
    \label{tab:dnsmos}
\end{table*} 

\subsection{Dataset}

Since the original DNSMOS model was trained on a proprietary, undisclosed dataset, we trained and evaluated our models on the public DNS2020 dataset \cite{reddy_interspeech_2020}.
This is a synthetic dataset that consists of clean speech with added noise at various \glspl{snr}.
Due to the lack of \gls{mos} target labels for this dataset, we computed \gls{pesq} using its standard intrusive implementation.
Thus, we tasked our model with unintrusively estimating \gls{pesq} using only the degraded (i.e., noisy) signal, as stated in the problem formulation (Methods Sec.~\ref{sec:problem}).
To this end, we used the official DNS2020 generation scripts\footnote{\url{https://github.com/microsoft/DNS-Challenge/blob/interspeech2020/master/noisyspeech_synthesizer_singleprocess.py}} to derive a dataset consisting of 6,000 noisy--clean pairs of 30\,s each, for a total of 50\,h of audio.
Subsequently, we framed each pair into 9-s segments with a stride of 2\,s (corresponding to 10 data points per clip) and computed a \gls{pesq} label for each segment, after which we discarded the clean reference signals.
In total, this amounts to 150\,h of overlapping audio segments.
We held out a validation set consisting of 5\,\% of the original 30-s clips.
For testing, we used the synthetic reverberation-free test data that is provided in the DNS2020 dataset, consisting of 150 10-s samples.

\subsection{Training and Evaluation Metrics}
\label{sec:training_eval}

All models were trained using stochastic gradient descent with a batch size of 128 and a learning rate of $10^{-3}$ for 400 epochs using the Adam optimizer with standard parameters and the \gls{mse} loss.
We established a scheduling policy to decrease the learning rate by a factor of 0.9 after 5 epochs without improvements, as well as an early stopping policy to automatically interrupt the training after 25 epochs without improvements.
The training took about 9.5 hours for the baseline model and 12 hours for the binarized model on an NVIDIA Quadro RTX 5000 GPU.
In line with previous work on \gls{sqp}, we evaluated the performance of the models using \gls{mse} and the \gls{pcc} \cite{kumar_torchaudio-squim_2023, reddy_dnsmos_2021, fu_quality-net_2018}. 
\section{Results}
\begin{figure}[tbh]
    \centering
    \includegraphics[width=\columnwidth]{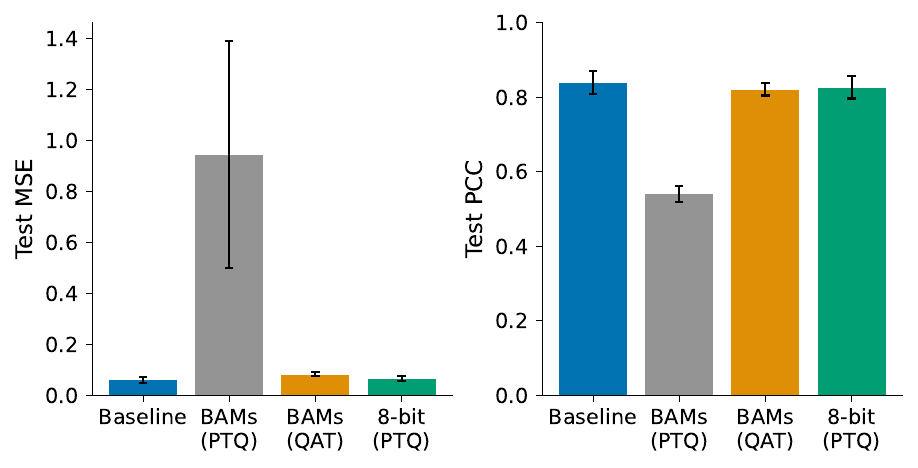}
    \caption{\textbf{Evaluation metrics.}
Test \gls{mse} and \gls{pcc} for the following different quantizations of DNSMOS:
        binary \gls{ptq} of activation maps (gray),
        \glspl{bam} with \gls{qat} (orange),
        and 8-bit \gls{ptq} of activations and weights (green).
The black lines indicate standard deviations from repeated instances of training (n=4).
    }
    \label{fig:metrics}
\end{figure}

To increase the computational efficiency of \gls{sqp}, we sought to investigate the performance impact of \glspl{bam} in the convolutional layers of the DNSMOS model.
First, we trained and evaluated the full-precision model (Table~\ref{tab:dnsmos}), thereby establishing a baseline for comparison.
Given the high number of convolutional operations in DNSMOS, binarizing the activation maps affects virtually all $2.3 \times 10^6$ activations, and replaces $171 \times 10^6$ of the $188 \times 10^6$ multiplications of the model (Table~\ref{tab:dnsmos}) with multiplications in which one of the factors is binary.
With suitable soft- and hardware support, such int1 multiplications have the potential to significantly reduce the computational and memory demands of the model.
We first measured the impact of post-training binarization of activation maps in the trained baseline model by simply replacing the activations in its convolutional layers with the Heaviside function.
This change resulted in a poorly performing model, with a reduction of test \gls{pcc} from $0.84 \pm 0.03$ to $0.54 \pm 0.02$, see Fig.~\ref{fig:metrics}.
Thus, binarization of the activation maps constitutes too drastic a precision reduction for performance to be maintained without employing some form of \gls{qat}.

\begin{figure}[tbh]
    \centering
    \includegraphics[width=\columnwidth]{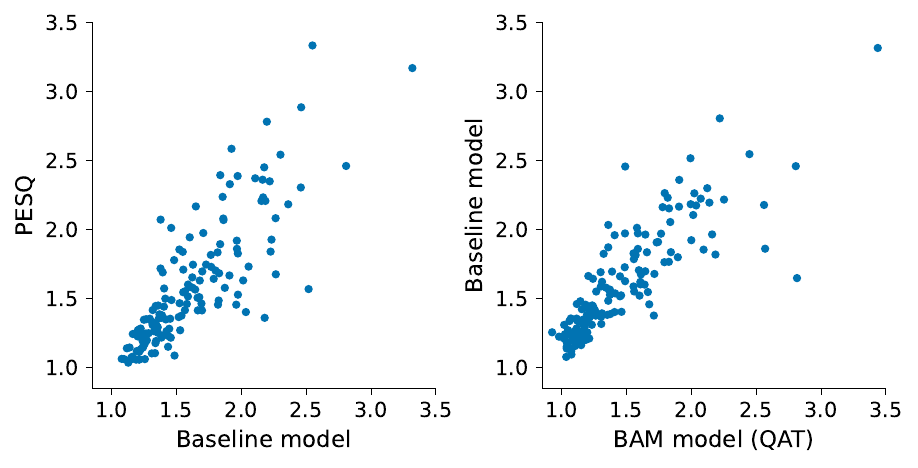}
    \caption{\textbf{Baseline model predictions vs.\ \gls{pesq} and \gls{bam} vs.\ baseline model predictions on the test set.}
    }
    \label{fig:preds_targets}
\end{figure}

To investigate the performance benefit of \gls{qat} on the binarized model, we conducted the binarization as before and trained the model from scratch using surrogate gradients (Methods Sec.~\ref{sec:surrogate_gradient}).
We first performed a coarse grid search to determine the optimal value of the steepness parameter, $\beta$.
We determined the optimal value $\beta=5$ on held-out data and used this value for all following experiments.
In contrast to the post-training binarization above, we found that \gls{qat} rescued the performance of the model, resulting in performance levels that were not significantly different from the baseline model, with $\mathrm{PCC}_\mathrm{BAM-QAT}=0.82 \pm 0.02$ and $\mathrm{PCC}_\mathrm{base}=0.84 \pm 0.03$, respectively, on the test set (Figs.~\ref{fig:metrics}~and~\ref{fig:preds_targets}).

Next, we tried binarizing the weights in addition to the activations, using \gls{qat} for training the model with fully binarized convolutions from scratch (Methods Sec.~\ref{sec:quantization}).
This change, however, resulted in substantially worse performance with a test \gls{pcc} of $0.63 \pm 0.05$.
We wondered whether there was some middle ground.
To test this idea, we next combined 8-bit \gls{ptq} of the weights, using PyTorch standard quantization mechanisms, with the \gls{bam} model trained using \gls{qat}.
This combination resulted in task performance of $0.81 \pm 0.01$ test \gls{pcc}, on par with the baseline.
At the same time, the binary mixed-precision model reduces the required memory per input sample during inference approximately 25-fold, from 9.66\,MB to 0.39\,MB, assuming a theoretical 1-bit representation of the \glspl{bam} and an 8-bit representation of the input spectrogram.
In conclusion, among the combinations we tested, a mixed-precision approach with \glspl{bam} and quantized weights seems most promising.

We wondered whether the above changes would already result in measurable computational gains, despite the fact that current deep learning libraries do not support mixed-precision quantization out of the box.
To test this idea, we used off-the-shelf 8-bit \gls{ptq} in PyTorch to quantize the above model (Methods Sec.~\ref{sec:quantization}).
Hence, in this setup, the 1-bit activations were wastefully represented with 8-bit tensors.
We measured the inference wall-clock time of the 8-bit model running on a CPU, as PyTorch does not support running quantized models on GPUs.
We found that the 8-bit quantized model showed reduced inference time by 56.3\,\% at comparable performance of $0.83 \pm 0.03$ test \gls{pcc} to the baseline model.
Thus, quantization has a measurable effect on inference time.
However, this reduction by approximately $2\times$ is only useful as a proxy, as the underlying implementation does not reap the performance gain that could be achieved by exploiting the int8~$\times$~int1 mixed-precision quantization.
Taking full advantage of this reduction will require special software and hardware support. 
\section{Discussion}
Here, we investigated how to make \gls{cnn}-based \gls{sqp} models more efficient through quantization.
To that end, we modified the DNSMOS architecture with \glspl{bam} and 8-bit \gls{ptq} of weights.
We found that our mixed-precision quantization scheme did not lead to significant decreases in task performance, while offering substantial memory and computational gains.
While we could not test how this extreme form of quantization translated into performance gains due to lack of suitable software support, we did measure a $2\times$ speed-up of inference time on CPU when we used off-the-shelf 8-bit \gls{ptq} of the whole model.
However, we expect that larger performance gains are possible on dedicated accelerators and software frameworks with full \gls{bam} support.

While these results are promising, the present study has several limitations.
First, since we derived our training data from a \gls{dns} dataset, it mostly comprises low-\gls{snr} samples, why our \gls{pesq} label distribution is considerably skewed towards low values.
Thus, for our system to be applicable in real-world scenarios, a training set with a more realistic distribution of ground-truth labels might prove necessary.
Second, our study is limited to a convolutional architecture and does not consider more sophisticated, and therefore more computationally demanding models \cite{kumar_torchaudio-squim_2023, fu_quality-net_2018}.
While such models may also benefit from \glspl{bam}, we deemed them less well suited for continuous resource-efficient \gls{sqp}.
Third, our weight quantization is limited to an 8-bit representation and only applied after training.
In future work, we intend to both investigate sub-8-bit quantization and systematically survey \gls{qat} for different weight-quantization levels.
Finally, our efficiency evaluation was limited by the lack of readily available hardware accelerators supporting \glspl{bam}.
An in-depth evaluation on mixed-precision hardware is left for future work.

In summary, our results indicate considerable efficiency gains in SQP by combining \glspl{bam} and non-binary weights on accelerators supporting mixed-precision arithmetic. 
\section{Acknowledgements}
This work was supported by
the Swiss National Science Foundation (Grant No.\ PCEFP3\_202981),
EU’s Horizon Europe Research and Innovation Programme (Grant Agreement No.\ 101070374, CONVOLVE) funded through SERI (Ref.\ 1131-52302),
and the Novartis Research Foundation. 
\bibliographystyle{IEEEtran}
\bibliography{references}

\begin{thebibliography}{10}
\providecommand{\url}[1]{#1}
\csname url@samestyle\endcsname
\providecommand{\newblock}{\relax}
\providecommand{\bibinfo}[2]{#2}
\providecommand{\BIBentrySTDinterwordspacing}{\spaceskip=0pt\relax}
\providecommand{\BIBentryALTinterwordstretchfactor}{4}
\providecommand{\BIBentryALTinterwordspacing}{\spaceskip=\fontdimen2\font plus
\BIBentryALTinterwordstretchfactor\fontdimen3\font minus \fontdimen4\font\relax}
\providecommand{\BIBforeignlanguage}[2]{{%
\expandafter\ifx\csname l@#1\endcsname\relax
\typeout{** WARNING: IEEEtran.bst: No hyphenation pattern has been}%
\typeout{** loaded for the language `#1'. Using the pattern for}%
\typeout{** the default language instead.}%
\else
\language=\csname l@#1\endcsname
\fi
#2}}
\providecommand{\BIBdecl}{\relax}
\BIBdecl

\bibitem{tzinis_sudo_2020}
\BIBentryALTinterwordspacing
E.~Tzinis, Z.~Wang, and P.~Smaragdis, ``Sudo rm -rf: {Efficient} {Networks} for {Universal} {Audio} {Source} {Separation},'' in \emph{2020 {IEEE} 30th {International} {Workshop} on {Machine} {Learning} for {Signal} {Processing} ({MLSP})}, Sep. 2020, pp. 1--6.
\BIBentrySTDinterwordspacing

\bibitem{bralios_latent_2023}
\BIBentryALTinterwordspacing
D.~Bralios, E.~Tzinis, G.~Wichern, P.~Smaragdis, and J.~L. Roux, ``Latent {Iterative} {Refinement} for {Modular} {Source} {Separation},'' in \emph{{ICASSP} 2023 - 2023 {IEEE} {International} {Conference} on {Acoustics}, {Speech} and {Signal} {Processing} ({ICASSP})}, Jun. 2023, pp. 1--5.
\BIBentrySTDinterwordspacing

\bibitem{fedorov_tinylstms_2020}
\BIBentryALTinterwordspacing
I.~Fedorov, M.~Stamenovic, C.~Jensen, L.-C. Yang, A.~Mandell, Y.~Gan, M.~Mattina, and P.~N. Whatmough, ``{TinyLSTMs}: {Efficient} {Neural} {Speech} {Enhancement} for {Hearing} {Aids},'' in \emph{Interspeech 2020}, Oct. 2020, pp. 4054--4058.
\BIBentrySTDinterwordspacing

\bibitem{rix_perceptual_2001}
\BIBentryALTinterwordspacing
A.~Rix, J.~Beerends, M.~Hollier, and A.~Hekstra, ``Perceptual evaluation of speech quality ({PESQ})-a new method for speech quality assessment of telephone networks and codecs,'' in \emph{2001 {IEEE} {International} {Conference} on {Acoustics}, {Speech}, and {Signal} {Processing}. {Proceedings} ({Cat}. {No}.{01CH37221})}, vol.~2, May 2001, pp. 749--752 vol.2.
\BIBentrySTDinterwordspacing

\bibitem{taal_algorithm_2011}
\BIBentryALTinterwordspacing
C.~H. Taal, R.~C. Hendriks, R.~Heusdens, and J.~Jensen, ``An {Algorithm} for {Intelligibility} {Prediction} of {Time}–{Frequency} {Weighted} {Noisy} {Speech},'' \emph{IEEE Transactions on Audio, Speech, and Language Processing}, vol.~19, no.~7, pp. 2125--2136, Sep. 2011.
\BIBentrySTDinterwordspacing

\bibitem{roux_sdr_2019}
\BIBentryALTinterwordspacing
J.~L. Roux, S.~Wisdom, H.~Erdogan, and J.~R. Hershey, ``{SDR} – {Half}-baked or {Well} {Done}?'' in \emph{{ICASSP} 2019 - 2019 {IEEE} {International} {Conference} on {Acoustics}, {Speech} and {Signal} {Processing} ({ICASSP})}, May 2019, pp. 626--630.
\BIBentrySTDinterwordspacing

\bibitem{reddy_scalable_2019}
C.~K. Reddy, E.~Beyrami, J.~Pool, R.~Cutler, S.~Srinivasan, and J.~Gehrke, ``A scalable noisy speech dataset and online subjective test framework,'' \emph{Proc. Interspeech 2019}, pp. 1816--1820, 2019.

\bibitem{fu_quality-net_2018}
\BIBentryALTinterwordspacing
S.-w. Fu, Y.~Tsao, H.-T. Hwang, and H.-M. Wang, ``\BIBforeignlanguage{en}{Quality-{Net}: {An} {End}-to-{End} {Non}-intrusive {Speech} {Quality} {Assessment} {Model} {Based} on {BLSTM}},'' in \emph{\BIBforeignlanguage{en}{Interspeech 2018}}.\hskip 1em plus 0.5em minus 0.4em\relax ISCA, Sep. 2018, pp. 1873--1877.
\BIBentrySTDinterwordspacing

\bibitem{jayesh_transformer_2022}
\BIBentryALTinterwordspacing
M.~K. Jayesh, M.~Sharma, P.~Vonteddu, M.~A.~B. Shaik, and S.~Ganapathy, ``\BIBforeignlanguage{en}{Transformer {Networks} for {Non}-{Intrusive} {Speech} {Quality} {Prediction}},'' in \emph{\BIBforeignlanguage{en}{Interspeech 2022}}.\hskip 1em plus 0.5em minus 0.4em\relax ISCA, Sep. 2022, pp. 4078--4082.
\BIBentrySTDinterwordspacing

\bibitem{yu_metricnet_2021}
\BIBentryALTinterwordspacing
M.~Yu, C.~Zhang, Y.~Xu, S.-X. Zhang, and D.~Yu, ``{MetricNet}: {Towards} {Improved} {Modeling} {For} {Non}-{Intrusive} {Speech} {Quality} {Assessment},'' in \emph{Proc. {Interspeech} 2021}, 2021, pp. 2142--2146.
\BIBentrySTDinterwordspacing

\bibitem{zezario_stoi-net_2020}
\BIBentryALTinterwordspacing
R.~E. Zezario, S.-W. Fu, C.-S. Fuh, Y.~Tsao, and H.-M. Wang, ``{STOI}-{Net}: {A} {Deep} {Learning} based {Non}-{Intrusive} {Speech} {Intelligibility} {Assessment} {Model},'' in \emph{2020 {Asia}-{Pacific} {Signal} and {Information} {Processing} {Association} {Annual} {Summit} and {Conference} ({APSIPA} {ASC})}, Dec. 2020, pp. 482--486.
\BIBentrySTDinterwordspacing

\bibitem{kumar_torchaudio-squim_2023}
\BIBentryALTinterwordspacing
A.~Kumar, K.~Tan, Z.~Ni, P.~Manocha, X.~Zhang, E.~Henderson, and B.~Xu, ``Torchaudio-{Squim}: {Reference}-{Less} {Speech} {Quality} and {Intelligibility} {Measures} in {Torchaudio},'' in \emph{{ICASSP} 2023 - 2023 {IEEE} {International} {Conference} on {Acoustics}, {Speech} and {Signal} {Processing} ({ICASSP})}, Jun. 2023, pp. 1--5.
\BIBentrySTDinterwordspacing

\bibitem{catellier_wideband_2023}
\BIBentryALTinterwordspacing
A.~A. Catellier and S.~D. Voran, ``\BIBforeignlanguage{en}{Wideband {Audio} {Waveform} {Evaluation} {Networks}: {Efficient}, {Accurate} {Estimation} of {Speech} {Qualities}},'' \emph{\BIBforeignlanguage{en}{IEEE Access}}, vol.~11, pp. 125\,576--125\,592, 2023.
\BIBentrySTDinterwordspacing

\bibitem{zezario_deep_2023}
\BIBentryALTinterwordspacing
R.~E. Zezario, S.-W. Fu, F.~Chen, C.-S. Fuh, H.-M. Wang, and Y.~Tsao, ``Deep {Learning}-{Based} {Non}-{Intrusive} {Multi}-{Objective} {Speech} {Assessment} {Model} {With} {Cross}-{Domain} {Features},'' \emph{IEEE/ACM Transactions on Audio, Speech, and Language Processing}, vol.~31, pp. 54--70, 2023.
\BIBentrySTDinterwordspacing

\bibitem{reddy_dnsmos_2021}
\BIBentryALTinterwordspacing
C.~K.~A. Reddy, V.~Gopal, and R.~Cutler, ``Dnsmos: {A} {Non}-{Intrusive} {Perceptual} {Objective} {Speech} {Quality} {Metric} to {Evaluate} {Noise} {Suppressors},'' in \emph{{ICASSP} 2021 - 2021 {IEEE} {International} {Conference} on {Acoustics}, {Speech} and {Signal} {Processing} ({ICASSP})}, Jun. 2021, pp. 6493--6497.
\BIBentrySTDinterwordspacing

\bibitem{mittag_nisqa_2021}
\BIBentryALTinterwordspacing
G.~Mittag, B.~Naderi, A.~Chehadi, and S.~Möller, ``{NISQA}: {A} {Deep} {CNN}-{Self}-{Attention} {Model} for {Multidimensional} {Speech} {Quality} {Prediction} with {Crowdsourced} {Datasets},'' in \emph{Interspeech 2021}, Aug. 2021, pp. 2127--2131.
\BIBentrySTDinterwordspacing

\bibitem{manocha_noresqa_2021}
\BIBentryALTinterwordspacing
P.~Manocha, B.~Xu, and A.~Kumar, ``{NORESQA}: {A} {Framework} for {Speech} {Quality} {Assessment} using {Non}-{Matching} {References},'' in \emph{Advances in {Neural} {Information} {Processing} {Systems}}, vol.~34.\hskip 1em plus 0.5em minus 0.4em\relax Curran Associates, Inc., 2021, pp. 22\,363--22\,378.
\BIBentrySTDinterwordspacing

\bibitem{elminshawi_slim-tasnet_2023}
\BIBentryALTinterwordspacing
M.~Elminshawi, S.~R. Chetupalli, and E.~A.~P. Habets, ``\BIBforeignlanguage{en}{Slim-{Tasnet}: {A} {Slimmable} {Neural} {Network} for {Speech} {Separation}},'' in \emph{\BIBforeignlanguage{en}{2023 {IEEE} {Workshop} on {Applications} of {Signal} {Processing} to {Audio} and {Acoustics} ({WASPAA})}}.\hskip 1em plus 0.5em minus 0.4em\relax New Paltz, NY, USA: IEEE, Oct. 2023, pp. 1--5.
\BIBentrySTDinterwordspacing

\bibitem{miccini_dynamic_2023}
\BIBentryALTinterwordspacing
R.~Miccini, A.~Zniber, C.~Laroche, T.~Piechowiak, M.~Schoeberl, L.~Pezzarossa, O.~Karrakchou, J.~Sparsø, and M.~Ghogho, ``Dynamic {nsNET2}: {Efficient} {Deep} {Noise} {Suppression} with {Early} {Exiting},'' in \emph{2023 {IEEE} 33rd {International} {Workshop} on {Machine} {Learning} for {Signal} {Processing} ({MLSP})}, Sep. 2023, pp. 1--6.
\BIBentrySTDinterwordspacing

\bibitem{chen_dont_2021}
\BIBentryALTinterwordspacing
S.~Chen, Y.~Wu, Z.~Chen, T.~Yoshioka, S.~Liu, J.~Li, and X.~Yu, ``Don’t {Shoot} {Butterfly} with {Rifles}: {Multi}-{Channel} {Continuous} {Speech} {Separation} with {Early} {Exit} {Transformer},'' in \emph{{ICASSP} 2021 - 2021 {IEEE} {International} {Conference} on {Acoustics}, {Speech} and {Signal} {Processing} ({ICASSP})}, Jun. 2021, pp. 6139--6143.
\BIBentrySTDinterwordspacing

\bibitem{li_learning_2021}
\BIBentryALTinterwordspacing
A.~Li, C.~Zheng, L.~Zhang, and X.~Li, ``Learning to {Inference} with {Early} {Exit} in the {Progressive} {Speech} {Enhancement},'' in \emph{2021 29th {European} {Signal} {Processing} {Conference} ({EUSIPCO})}, Aug. 2021, pp. 466--470.
\BIBentrySTDinterwordspacing

\bibitem{sun_progressive_2020}
\BIBentryALTinterwordspacing
L.~Sun, J.~Du, X.~Zhang, T.~Gao, X.~Fang, and C.-H. Lee, ``Progressive {Multi}-{Target} {Network} {Based} {Speech} {Enhancement} with {Snr}-{Preselection} for {Robust} {Speaker} {Diarization},'' in \emph{{ICASSP} 2020 - 2020 {IEEE} {International} {Conference} on {Acoustics}, {Speech} and {Signal} {Processing} ({ICASSP})}, May 2020, pp. 7099--7103.
\BIBentrySTDinterwordspacing

\bibitem{bengio_estimating_2013}
\BIBentryALTinterwordspacing
Y.~Bengio, N.~Léonard, and A.~Courville, ``Estimating or {Propagating} {Gradients} {Through} {Stochastic} {Neurons} for {Conditional} {Computation},'' Aug. 2013, arXiv:1308.3432 [cs]. [Online]. Available: \url{http://arxiv.org/abs/1308.3432}
\BIBentrySTDinterwordspacing

\bibitem{ho_naaloss_2023}
\BIBentryALTinterwordspacing
K.-H. Ho, E.-L. Yu, J.-W. Hung, and B.~Chen, ``\BIBforeignlanguage{en}{{NAaLOSS}: {Rethinking} the {Objective} of {Speech} {Enhancement}},'' in \emph{\BIBforeignlanguage{en}{2023 {IEEE} 33rd {International} {Workshop} on {Machine} {Learning} for {Signal} {Processing} ({MLSP})}}.\hskip 1em plus 0.5em minus 0.4em\relax Rome, Italy: IEEE, Sep. 2023, pp. 1--6.
\BIBentrySTDinterwordspacing

\bibitem{courbariaux_binarized_2016}
\BIBentryALTinterwordspacing
M.~Courbariaux, I.~Hubara, D.~Soudry, R.~El-Yaniv, and Y.~Bengio, ``Binarized {Neural} {Networks}: {Training} {Deep} {Neural} {Networks} with {Weights} and {Activations} {Constrained} to +1 or -1,'' Mar. 2016, arXiv:1602.02830 [cs]. [Online]. Available: \url{http://arxiv.org/abs/1602.02830}
\BIBentrySTDinterwordspacing

\bibitem{neftci_surrogate_2019}
\BIBentryALTinterwordspacing
E.~O. Neftci, H.~Mostafa, and F.~Zenke, ``\BIBforeignlanguage{en}{Surrogate {Gradient} {Learning} in {Spiking} {Neural} {Networks}: {Bringing} the {Power} of {Gradient}-{Based} {Optimization} to {Spiking} {Neural} {Networks}},'' \emph{\BIBforeignlanguage{en}{IEEE Signal Processing Magazine}}, vol.~36, no.~6, pp. 51--63, Nov. 2019.
\BIBentrySTDinterwordspacing

\bibitem{jacob_quantization_2018}
\BIBentryALTinterwordspacing
B.~Jacob, S.~Kligys, B.~Chen, M.~Zhu, M.~Tang, A.~Howard, H.~Adam, and D.~Kalenichenko, ``Quantization and {Training} of {Neural} {Networks} for {Efficient} {Integer}-{Arithmetic}-{Only} {Inference},'' in \emph{Proceedings of the {IEEE} {Conference} on {Computer} {Vision} and {Pattern} {Recognition} ({CVPR})}, 2018, pp. 2704--2713.
\BIBentrySTDinterwordspacing

\bibitem{zenke_superspike_2018}
\BIBentryALTinterwordspacing
F.~Zenke and S.~Ganguli, ``{SuperSpike}: {Supervised} {Learning} in {Multilayer} {Spiking} {Neural} {Networks},'' \emph{Neural Computation}, vol.~30, no.~6, pp. 1514--1541, Jun. 2018.
\BIBentrySTDinterwordspacing

\bibitem{dubey_icassp_2024}
\BIBentryALTinterwordspacing
H.~Dubey, A.~Aazami, V.~Gopal, B.~Naderi, S.~Braun, R.~Cutler, A.~Ju, M.~Zohourian, M.~Tang, M.~Golestaneh, and R.~Aichner, ``{ICASSP} 2023 {Deep} {Noise} {Suppression} {Challenge},'' \emph{IEEE Open Journal of Signal Processing}, pp. 1--13, 2024.
\BIBentrySTDinterwordspacing

\bibitem{timcheck_intel_2023}
\BIBentryALTinterwordspacing
J.~Timcheck, S.~B. Shrestha, D.~B.~D. Rubin, A.~Kupryjanow, G.~Orchard, L.~Pindor, T.~Shea, and M.~Davies, ``\BIBforeignlanguage{en}{The {Intel} neuromorphic {DNS} challenge},'' \emph{\BIBforeignlanguage{en}{Neuromorphic Computing and Engineering}}, vol.~3, no.~3, p. 034005, Aug. 2023.
\BIBentrySTDinterwordspacing

\bibitem{Yep_torchinfo_2020}
\BIBentryALTinterwordspacing
T.~Yep, ``torchinfo,'' Mar. 2020. [Online]. Available: \url{https://github.com/TylerYep/torchinfo}
\BIBentrySTDinterwordspacing

\bibitem{reddy_interspeech_2020}
\BIBentryALTinterwordspacing
C.~K. Reddy, V.~Gopal, R.~Cutler, E.~Beyrami, R.~Cheng, H.~Dubey, S.~Matusevych, R.~Aichner, A.~Aazami, S.~Braun, P.~Rana, S.~Srinivasan, and J.~Gehrke, ``\BIBforeignlanguage{en}{The {INTERSPEECH} 2020 {Deep} {Noise} {Suppression} {Challenge}: {Datasets}, {Subjective} {Testing} {Framework}, and {Challenge} {Results}},'' in \emph{\BIBforeignlanguage{en}{Interspeech 2020}}.\hskip 1em plus 0.5em minus 0.4em\relax ISCA, Oct. 2020, pp. 2492--2496.
\BIBentrySTDinterwordspacing

\end{thebibliography}

\end{document}